\documentclass[10pt]{article}
\usepackage{amssymb,amsmath,amsthm}
\usepackage[pdftex]{graphicx}
\usepackage{url}
\usepackage{natbib}

\newcommand\acc{\^}

\newcommand\be{$$}
\newcommand\ee{$$}
\newcommand\ben{\begin{equation}}
\newcommand\een{\end{equation}}
\newcommand\bea{\begin{eqnarray*}}
\newcommand\eea{\end{eqnarray*}}
\newcommand\bean{\begin{eqnarray}}
\newcommand\eean{\end{eqnarray}}

\begin{document}

\section*{Self-Financing Trading and the It\acc{o}-D\"{o}blin Lemma\footnote{\bf The views expressed are those of the authors only, no other representation should be attributed.   Not guaranteed fit for any purpose.  Use at your own risk.}\\\large Chris Kenyon\footnote{Contact: chris.kenyon@lloydsbanking.com} and Andrew Green\footnote{Contact: andrew.green2@lloydsbanking.com}}

The objective of the note is to remind readers on how self-financing works in Quantitative Finance.  The authors have observed continuing uncertainty on this issue which may be because it lies exactly at the intersection of stochastic calculus and finance.  The concept of a self-financing trading strategy was originally, and carefully, introduced in \cite{Harrison1979a} and expanded very generally in \cite{Harrison1981a}.

\paragraph{The Issue}   The value $Y_t$ of a portfolio (using notation as \cite{Duffie2001a}) composed of stock $S_t$ and bond $\beta_t$ with holding $a_t$ and $b_t$ can be written (Equation 14 on page 90):
\be
Y_t = a_t S_t + b_t \beta_t 
\ee
the change in portfolio value, or {\it gain process} is given as (Equation 15 on page 90):
\be
dY_t = a_t dS_t + b_t d\beta_t
\ee
Clearly, if $a_t$ is a delta hedge, i.e. a function of $S_t$, then applying the It\acc{o}-D\"{o}blin Lemma to the  equation for $Y_t$ would give:
\be
dY_t = a_t dS_t + S_t da_t + da_t dS_t + b_t d\beta_t + \beta_t d b_t + db_t d\beta_t
\ee
and the $\beta_t d b_t + db_t d\beta_t$ terms are also simply a mathematical consequence of applying the Lemma.
So has Prof Duffie made a mistake that is still there in the 3rd edition of his text?  This is the crux of this issue at the intersection between stochastic calculus (the It\acc{o}-D\"{o}blin Lemma) and finance (Duffie's equation 15), i.e. the concept of a self-financing portfolio.

\paragraph{The Resolution} is simply the {\it definitions} in \cite{Harrison1979a,Harrison1981a} and reproduced in \cite{Duffie2001a}  that a self-financing portfolio follows (page 89):
\ben
a_t S_t + b_t \beta_t = a_0 S_0 + b_0 \beta_0 + \int_0^t a_u dS_u + \int_0^t b_t d\beta_u \label{e:d1}
\een
or
\ben
d(a_t S_t + b_t \beta_t) = a_t dS_t + b_t d\beta_t \label{e:d2}
\een
What this says is that the only change in portfolio value comes from the value of the stock and bond (or cash account), whatever the trading strategy.  The trading strategy can move value between the stock and cash accounts but not create or destroy value.  If this were not true then the basic result that all self-financing portfolios have the same rate of return in the risk-neutral measure would be false (\cite{Harrison1981a}).

Basically {\it by definition of self-financing}  the only change in portfolio value comes from the value of the underlyings (the gain process).  An additional self-financing\footnote{Note that ``self-financing condition'' or equation is applied to different pieces of this setup by different authors.} equation is {\it implied}, here $S_t da_t + da_t dS_t + \beta_t d b_t + db_t d\beta_t \equiv 0$, but it adds nothing since it is simply a direct consequence of the {\it definition} of self-financing. However, it is irrelevant because it is the definition that drives the theory.

\paragraph{Discussion} 
In short, you cannot apply the It\acc{o}-D\"{o}blin Lemma to a portfolio's value expressed in terms of its underlyings and get its gain process. This is by definition (\cite{Harrison1979a,Harrison1981a}).  

The definition of a self-financing trading strategy chosen by \cite{Harrison1979a,Harrison1981a} means that continuous time works like a limit of discrete time trading. Trading strategies are predictable (no use of the future) and have additional technical limits (e.g. quadratic bounds), consistent with the It\acc{o}-D\"{o}blin Lemma, that rule out things like doubling strategies and other unwanted arbitrage mechanisms.

Mathematically it is possible to chose other definitions of self-financing from those chosen by \cite{Harrison1979a,Harrison1981a}.  Then you have a different theory, and one that is not what is currently accepted, and been found useful over the last thirty or so years in Quantitative Finance.  The key point delivered by the {\it definition} is that all self-financing portfolio provide the same rate of return in the risk neutral measure.  If the trading strategy could change the rate of return then the theory would be broken as arbitrage opportunities would be immediate.  Hence we see that the current {\it definition} of self-financing, that portfolio values changes only through its underlyings, is appropriate for Quantitative Finance.  Recent modifications \cite{Kenyon2014b} build on this framework, they do not contradict it.

\small
\paragraph{Acknowledgements}  The authors would like to thank Prof Darrell Duffie for useful pointers.  Any errors remain their own.
\bibliographystyle{plainnat}
\bibliography{kenyon_general}
\end{document}